\documentstyle[11pt, a4wide, epsf]{article} % ]{crckapb}
\begin{document}
\centerline{\bf X--RAY EMISSION IN GAMMA-RAY BLAZARS}
\vskip 0.5 true cm
\centerline{Gabriele Ghisellini}
\centerline{\it Osservatorio di Milano, V. Bianchi, 46, Merate 
(Lecco), Italy}
\noindent

%\vskip 0.5 true cm
%\noindent 
%{\bf Abstract}
%Although the $\gamma$--ray emission in blazars dominates the power output, 
%there are crucial informations carried by the X--rays.
%Indeed, their $paucity$, together with the short variability timescales 
%observed both at X and $\gamma$--ray energies constrains the location, 
%along the jet, of the active region responsible for most of the emission.
%Furthermore, the knowledge of the X--ray flux and spectrum helps in 
%defining the overall spectral energy distribution (SED) of blazars, 
%allowing to discriminate between models.

\vskip 0.3 true cm
\noindent
{\bf 1. Low entropy inner jets}
\vskip 0.2 true cm

\noindent
As shown by Maraschi (this volume) the overall spectral energy distribution 
(SED) of the blazars detected by EGRET shows two peaks, in a 
$\nu$--$\nu F(\nu)$ plot.
With few exceptions, the minimum between these two peaks occurs in the 
X--ray range.
We can exploit this observational fact to constrain the models proposed 
to account for the SED of blazars, and to reach an important conclusion: 
the conversion of the primary power carried by the jet into radiation 
occurs primarily at some distance from the central powerhouse.

Assume in fact that part of the $\gamma$--ray radiation is absorbed
by $\gamma$--$\gamma$ collisions (Blandford \& Levinson 1995).
%Blandford 1993, 
This implies that, in the comoving frame of the jet/blob emitting the high 
energy radiation, there is a sufficient amount of target X--rays. 
The pairs created in this way are relativistic, and can emit at lower 
frequencies, or escape, if their cooling time is sufficiently long.
In the first case, all the absorbed power in the $\gamma$--ray band
reappears at lower energies, namely the X--ray band.
Therefore it is inevitable to predict that the X--ray luminosity
should be of the same order of the $\gamma$--ray luminosity.

A way out of this is to assume that the cooling time of the pairs is longer
than their escape time.
In this case only a fraction of the power absorbed in the $\gamma$--ray band
is reprocessed into radiation of softer energy, mainly X--rays.
This model then requires that there are sufficient X--rays to 
absorb the $\gamma$--rays, but not enough photons for Compton cooling.
In principle, this is possible, because the scattering between pairs and 
X--rays occurs in the inefficient Klein Nishina regime, but it is highly 
unlikely, because the X--ray emission should always be accompanied by at 
least a comparable amount of optical UV radiation.
Requiring efficient absorption of $\gamma$--rays, therefore, implies short 
cooling times of the pairs, resulting in an excess of X--ray emission
(Ghisellini \& Madau 1996).
Since this is not observed, we conclude that:

\noindent
$\bullet$ The $\gamma$--ray emitting region is transparent.

\noindent
$\bullet$ In order to be transparent, it must be at some distance
from the (X--ray emitting) accretion disk.
On the other hand, the short variability timescales observed
at high energies limit the dimensions of the source, and hence
its location in the jet. 
By combining both limits, we can derive a typical distance
at which dissipation occurs, of few hundreds Schwarzschild radii.

\noindent
$\bullet$ In the inner part of the jet, energy must be transported
efficiently, without dissipation.
Possibilities are:
i) cold (in the comoving frame) protons with bulk Lorentz factor $\Gamma$;
ii) Poynting flux.
The kinetic luminosity carried by the protons is
$$
L_k\, =\, \pi r^2 n_p\Gamma\beta m_pc^3\, \to
\tau_p\, \sim \, 5\times 10^{-2} {\left(L_{k,46} \over \Gamma_1 r_{14}
\right)}
$$
where $r$ is the cross sectional radius of the jet,
$\tau_p=\sigma_T n_p r$, $\Gamma=10\Gamma_1$ and
$L_k=10^{46}L_{k,46}$ erg/s.
As Sikora (this volume) points out,
cold electrons of the same optical depth would produce an observable
bump at $\sim$1 keV by scattering ambient UV photons coming
from the accretion disk and from the broad line region.
This bump is not observed,
{\it leaving Poynting flux as the only viable possibility}.

\vfill
\eject
% \vskip 0.3 true cm
\noindent
{\bf 2. One or two electron populations?}
\vskip 0.2 true cm

\noindent
Mannheim (1993, see also this volume) suggested that the overall emission 
in blazars can be the result of two different electron populations, 
both emitting by the synchrotron mechanism.
Radiation losses can in fact limit the maximum attainable Lorentz 
factors of accelerated electrons, but are not important for protons,
which can therefore reach ultra--relativistic energies.
By interacting with the photons produced by the `primary' electrons,
an $e^\pm$ pair cascade develops, with the created pairs reaching 
Lorentz factors up to $\gamma\sim 10^{10}$ or more.
The first population of electrons should be responsible for the first 
(low energy) synchrotron peak, while the pairs produced by the protons 
produce the high energy peak.
In this scenario the location of the two peaks are somewhat unrelated, 
corresponding to the maximum energy of the $two$ electron populations.

In synchrotron--Compton models, instead, a single electron
population is responsible for both peaks of the SED.
Even if we do not know (yet) the origin of the seed photons
to be upscatterd at high energies, both peaks are produced
by the same electrons, of Lorentz factor, say, $\gamma_b$.
A distribution of $\gamma_b$--values results in a corresponding
distribution of peak energies of the synchrotron ($\nu_S$) and
Compton ($\nu_C$) components.
Therefore sources whose synchrotron component peaks in the far IR
should have `Compton' peaks in the MeV range, while sources with
$\nu_S$ in the soft X--ray band should show a Compton peak
in the GeV to TeV region.
The three sources detected so far in the TeV band by Whipple
indeed confirm this scenario.

Other evidence come from the (anti)-correlation found by Comastri
et al. (1996) between $\alpha_{ro}$, the index connecting the
5 GHz flux with the optical (V band) flux, and $\alpha_{x\gamma}$,
the index between 1 keV and 100 MeV.
The found anticorrelation can be easily explained by a distribution
of $\gamma_b$. 
If it is large, as in the Whipple detected BL Lacs, then $\nu_S$
is above the optical band, and the radio to optical index is flat.
At the same time the X--ray flux is large with respect to the 
100 MeV emission, since the synchrotron emission extends to the X--rays
(it peaks there), while the $\gamma$--ray spectrum, at
100 MeV, is still rising.
The reverse (flat $\alpha_{x\gamma}$ when $\alpha_{ro}$ is steep)
corresponds instead to a small value of $\gamma_b$.

A distribution of $\gamma_b$ can also explain the correlation found
by Comastri et al. (1996)  
between $\alpha_X$ and $\alpha_\gamma$.
If $\alpha_x$ is steep, the X--rays are produced in the tail of
the synchrotront component, which therefore should have
a large $\nu_S$ and a large $\gamma_b$.
This leads to a Compton peak in the GeV to TeV band, and then to 
a flat value of $\alpha_\gamma$ in the EGRET band.

We therefore conclude that it is easier to explain these correlations 
in the framework of models involving only one electron popoulation.
{\it This implies that by monitoring the synchrotron
emission around $\nu_S$ we monitor the same electrons
producing the Compton flux close to $\nu_C$.}

%\vskip 0.5 true cm
%\centerline{\bf 3. X--rays and the kinetic luminosity of the jet}
%\vskip 0.3 true cm

%\noindent
%By applying the simple, one zone SSC model to the radio core,
%it is possible to derive a limit on the amount of electrons
%responsible for the radio and the X--ray emission of the core itself.
%Note that this limit is, at least in principle, independent of the degree 
%of beaming, since it can be derived simply making the ratio of the
%self-Compton and synchrotron fluxes at the same frequency.
%Then, by comparing the predicted (from radio data) and the observed 
%X--ray flux one also derives a lower limit on the Doppler factor $\delta$.
%Another way to derive it is to assume that the X--ray and the $\gamma$--ray
%are cospatial, and to require transparency (Dondi \& Ghisellini, 1995).
%It is then possible to obtain all the quantities necessary to derive the 
%kinetic luminosity: the cross sectional radius of the jet (directly from 
%VLBI angular sizes), the electron density and the $\Gamma\sim \delta$ 
%factor.
%In this way, Celotti \& Fabian (1993) derived the kinetic luminosities
%for a sample of radio sources, finding that it is of the same
%order of the power required by the extended lobes to exist,
%much greater than the intrinsic emitted power.
%{\it This implies that the jet is a inefficient radiator}.

\vskip 0.3 true cm
\noindent
{\bf 3. Fast variability of X--rays}
\vskip 0.2 true cm

\noindent
In the X--ray band we probably reach the best trade off between
fast variability and amount of photons needed to detect it.
The light curve of the BL Lac object PKS 2155--304 (which was
observed simultaneously with ASCA, EUVE and IUE; see Maraschi,
this volume) is probably one of the best examples, together
with the recent results about Mkn 421 observed by ASCA, 
in which delays between the hard and soft X--ray band have
been observed (Takahashi et al. 1996).
The symmetry (i.e. equal rise and decay times) of the flare of PKS 2155--304 
in the ASCA band is highly indicative that the involved timescale 
is neither connected with the acceleration nor the cooling timescales,
but it is instead related to $R/c$.
This immediately implies that:

\noindent
$\bullet$ The cooling timescale is shorter than $R/c$. 
If the cooling is radiative, as it is likely especially at high 
energies, then we can set a limit on the magnetic field
(see, e.g. Massaro et al. 1996).

\noindent
$\bullet$ The electron distribution responsible for the emission
is evolving rapidly, faster than $R/c$.
{\it The observer will see a convolution of different spectra, each
produced in a different region of the source}.
Initially we see only the emission by fresh electrons located in the
slice nearest to us. After a time $R/c$ we see the entire source:
the back of it with fresh electrons, and the front of it with older
electrons. In this way Chiaberge \& Ghisellini (1996) could
explain the time-delay between hard and soft X--ray
observed in Mkn 421, as illustrated in Fig. 2.

\vskip 0.3 true cm
\noindent
{\bf 4. Very flat X--ray indices?}
\vskip 0.2 true cm
\noindent
The so called `MeV blazars' are characterized by a steep 
($\alpha_\gamma >1$) $\gamma$-ray spectrum, suggesting that the
peak of their Compton component lies in the MeV range.
This has been directly confirmed in some of these sources by COMPTEL data.
Although not simultaneous, the X--ray data in these sources
indicate a very large $\gamma$ to X--ray flux ratio, and
hence a very flat X--ray spectral index, even flatter than $\alpha_x=0.5$.
If confirmed (by, e.g. SAX), this could give valuable information
on the origin of the high energy emission.
There can be several possible alternatives:

\noindent
$\bullet$ Sikora et al. (1996) found that MeV blazars are the most difficult
to be explained both in the SSC and in the external photon scenario.
They suggest that in these sources the electron distribution is not the 
result of injection and cooling or escape, but it reaches a steady 
state through the competition of reacceleration and cooling.
This leads to a particular electron energy where heating and
cooling balance, which can be be of the order of 100 MeV.
Such a peaked electron distribution, via Compon scattering,
produces a very flat spectrum [the limit being $F(\nu)
\propto \nu$].

\noindent
$\bullet$ A flat electron distribution can also be the result of 
incomplete cooling.
In fact, assume to continuously inject a power law distribution 
of electrons $Q(\gamma)\propto \gamma^{-s}$.
If the cooling time at all energies is shorter than the escape time, 
the equilibrium electron distribution is $N(\gamma) = \int_\gamma 
Q(\gamma)d\gamma / \dot \gamma$, where $\dot\gamma\propto \gamma^2$ 
is the cooling rate for synchrotron and Compton losses.
The flattest distribution is  $N(\gamma)\propto \gamma^{-2}$, corresponding 
to $F(\nu)\propto \nu^{-0.5}$.
However, if the cooling time is longer than the escape time, then 
$N(\gamma)\sim Q(\gamma) t_{esc}$, and in this case  one can obtain 
a spectral index $\alpha$ flatter than 0.5.
A break occurs for $t_{cool}(\gamma) = t_{escape}$.
If Compton cooling is dominant, then $\gamma_b\sim 3\pi/\ell$ where 
$\ell = L\sigma_T/(Rmc^3)$ is the compactness as seen in the comoving 
frame of the blob, which is the sum of the locally produced synchrotron 
radiation and the externally produced (and Doppler boosted) emission.

\noindent
$\bullet$ Assume that the high energy emission is the result of Compton 
scattering with external photons.
In the comoving (primed) frame of the blob, these photons are seen blushifted
by $\sim \Gamma$. 
Assume also that their spectrum is not monochromatic, but it 
extends between $\nu^{\prime}_1$ and $\nu^{\prime}_2$.
Above $\nu_2^\prime$, all incident photons are used to form the 
Comptonized spectrum, but for $\nu<\nu^{\prime}_2$ only part of 
the incident photons can be used.
Then the Comptonized spectrum shows a break at $\nu_2^\prime$,
being flatter below.
In the observing frame, this break is visible, because the 
Comptonized spectrum is Doppler boosted and blueshifted, while 
the incident spectrum (with respect to what the blob sees), is 
redshifted by a factor $\sim \Gamma$.

To illustrate the latter case, I have tried to model the overall
spectrum of the blazar 0202+149, which has the flattest X--ray slope 
(as determined by ROSAT) of the $\gamma$--bright blazars analized
by Comastri et al. (1996).
As can be seen, the X--ray spectrum has a complex shape, which is
the result of SSC and external Compton contributions.
The former contributes below 0.1 keV. 
Above this energy, Compton scattering with radiation produced
externally to the jet dominates, producing a very flat
spectrum between 1 and 10 keV: in this band
the power rises steeply with frequency because of the 
increasing numer of photons that can be used for the scattering
process. 
Above 10 keV all seed photons are used, and the spectrum has
the canonical 0.5 slope.

\noindent
$\bullet$  Cold (in the comoving frame) electrons partecipating to the bulk
motion can contribute to the X--ray emission at 
$\nu\sim \Gamma^2 \nu_{UV}\sim$ 1 keV, producing the ``Sikora bump"
(Sikora et al. 1996, see also this volume).
Its amplitude depends on the scattering optical depth of these 
electrons.
In this way Sikora et al. (1996) can constrain the amount
of cold electrons and $e^\pm$ allowed to be present in the inner jet, since
the ``Sikora bump" is not (yet) observed.
Note that even more stringent constraints can be derived
in the very inner part of the jet, where the radiation coming
directly from the accretion disk (neglected in Sikora et al. 1996)
is likely to dominate the radiation energy density as seen by the jet
(Ghisellini \& Madau, 1996).

\begin{figure}
\epsfxsize = 6 in 
\epsffile{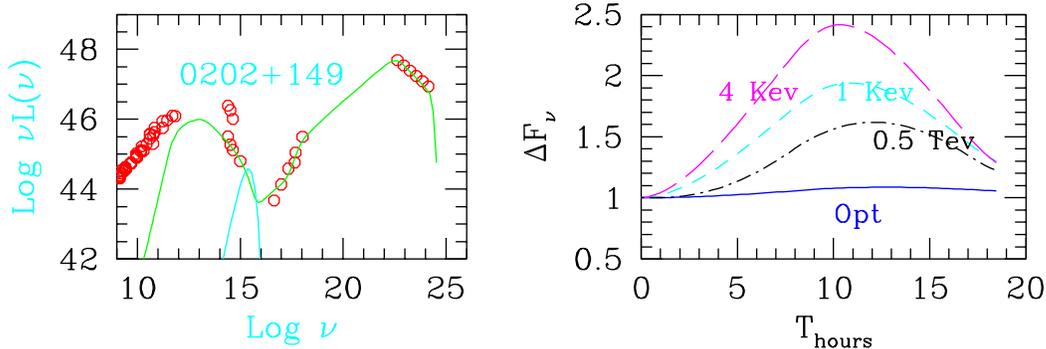}
\vskip -10.6 true cm
\caption [] {The overall spectrum of 0202+149 fitted by
the external Compton model. The external photons are assumed
to be distributed as a diluted blackbody. The fast rise in the 
soft X--ray band is due to the increasing (with energy)
number of incident photons available for the inverse Compton
process. 
Fig. 2: Time lags between hard and soft X--rays, TeV and
optical emission in Mkn 421. High energy electrons are
injected for a time $\sim R/c$, with $R=10^{16}$ cm.
}
\end{figure}

\vskip 0.3 true cm
\noindent
{\bf References}
\parindent=0 pt
\everypar={\hangindent=2.6pc}
\vskip 0.2 true cm

%Blandford, R. D., 1993, in {\it 1st Compton Gamma-Ray Observatory Symposium},
%AIP Conference Proc, 280, ed. M.Friedlander et al. (New York: AIP), 533.
 
Blandford, R.D. \& Levinson, A. 1995, ApJ, 441, 79

% Celotti A., Fabian A.C., 1993, MNRAS, 264, 228

Chiaberge, M. \& Ghisellini, G., 1996, in ``Second italian meeting 
on AGNs", in press.

Comastri, A., Fossati, G., Ghisellini, G. \& Molendi, S., 1996, submitted
to ApJ

% Dermer, C.D., 1995, ApJ, 446, L63

% Dermer, C.D., \& Schlickeiser, 1993, ApJ, 416, 458.

Dondi, L., \& Ghisellini, G., 1995, MNRAS, 273, 583.

Ghisellini G. \& Madau P., 1996, MNRAS, 280, 67

Mannheim, K., 1993, A\& A 269, 67

%Maraschi, L., Ghisellini, G. \& Celotti, A. 1994, in {\it Multi-wavelength 
%continuum Emission of AGN} IAU Symp. N. 159, Eds T.J.L. Courvoisier 
%\& A. Blecha (Kluwer) 233.

%Maraschi, L., Ghisellini, G. \& Celotti, A. 1992, ApJ, 397, L5

%Maraschi, L., et al. 1994, ApJ, 435, L91
 
%Maraschi, L., et al., 1994, ApJ, 435, 91.

%Marscher, A.P., \& Bloom S.D., 1994,  in {\it 2nd Compton Observatory 
%Symposium}, AIP Conf. Series, Eds. C.E. Fichtel et al. (New York: AIP).

Massaro, E. et al., 1996, A\&A, in press

% Sikora, M., Begelman, M.C. \& Rees, M.J. 1993, in Compton Gamma-ray
% Observatory, AIP Proceedings 280, eds. M. Friedlander, N. Geherels
% \& D.J. Macomb (New York: AIP), p. 598

% Sikora, M., Begelman, M.C. \& Rees, M.J. 1994, ApJ, 421, 153
 
% Sikora, M. 1994, ApJS, 90, 923.

Sikora, M., Madejski, G. \& Moderski, R., 1996, submitted to ApJ.

Takahashi, T. et al. 1996, ApJ in press

% von Montigny, C., et al. 1995, ApJ, 440, 525

\end{document}